\documentclass[final,3p,times,twocolumn]{elsarticle}
\pdfoutput=1 

\usepackage[utf8]{inputenc}
\usepackage{textcomp}
\usepackage{lineno}
\usepackage{color}

\usepackage[outdir=./]{epstopdf}
\usepackage{upgreek}
\usepackage{amsmath}
\usepackage{subcaption}
\usepackage{hyperref}
\begin{document}

\begin{frontmatter}
  \title{Beam test performance of the highly granular SiW-ECAL technological prototype for the ILC.}


  \author[kyushu]{K.\,Kawagoe}
  \author[kyushu]{Y.\,Miura}
  \author[kyushu]{I.\,Sekiya}
  \author[kyushu]{T.\,Suehara}
  \author[kyushu]{T.\,Yoshioka}
  
  \author[lal]{S.\,Bilokin\corref{cor_iphc}}
  \author[lal]{J.\,Bonis}
  \author[lal]{P.\,Cornebise}
  \author[lal]{A.\,Gallas}
  \author[lal]{\underline{A.\,Irles}\corref{cor1}}
  \ead{irles@lal.in2p3.fr}
  \author[lal]{R.\,P\"oschl}
  \author[lal]{F.\,Richard}
  \author[lal]{A.\,Thiebault}
  \author[lal]{D.\,Zerwas}

  \author[llr]{M.\,Anduze}
  \author[llr]{V.\,Balagura}
  \author[llr]{V.\,Boudry}
  \author[llr]{J-C.\,Brient}
  \author[llr]{E.\,Edy}
  \author[llr]{G.\,Fayolle}
  \author[llr]{M.\,Frotin}
  \author[llr]{F.\,Gastaldi}
  \author[llr]{R.\,Guillaumat}
  \author[llr]{A.\,Lobanov}
  \author[llr]{M.\,Louzir}
  \author[llr]{F.\,Magniette}
  \author[llr]{J.\,Nanni}
  \author[llr]{M.\,Rubio-Roy\corref{cor_iphc}}
  \author[llr]{K.\,Shpak}
  \author[llr]{H.\,Videau}
  \author[llr,ihep]{D.\,Yu}

  \author[omega]{S.\,Callier}
  \author[omega]{F.\,Dulucq}
  \author[omega]{Ch.\,de la Taille}
  \author[omega]{N.\,Seguin-Moreau}

  \author[lpnhe]{J.E.\,Augustin}
  \author[lpnhe]{R.\,Cornat}
  \author[lpnhe]{J.\,David}
  \author[lpnhe]{P.\,Ghislain}
  \author[lpnhe]{D.\,Lacour}
  \author[lpnhe]{L.\,Lavergne\corref{cor_iphc}}
  \author[lpnhe]{J.M.\, Parraud}

  \author[skku]{J.\,S.\,Chai}
  
  \author[kek]{D.\,Jeans}

  \address[kyushu]{Department of Physics and Research Center for Advanced Particle Physics, 
      Kyushu University, 744 Motooka, Nishi-ku, Fukuoka 819-0395, Japan }

  \address[lal]{Laboratoire de l'Acc\'elerateur Lin\'eaire (LAL), CNRS/IN2P3 et
      Universit\'e de Paris-Sud XI, Centre Scientifique d'Orsay B\^atiment 200, BP 34, F-91898 Orsay 
    CEDEX, France}

  \address[llr]{Laboratoire Leprince-Ringuet (LLR)
    -- \'{E}cole polytechnique, CNRS/IN2P3, F-91128 Palaiseau Cedex, France }
  \address[ihep]{Institute of High Energy Physics of Beijing (IHEP), 19 Yuquan Rd, Shijingshan Qu, Beijing Shi, China}

  \address[omega]{Laboratoire OMEGA -- \'{E}cole polytechnique-CNRS/IN2P3, F-91128 Palaiseau Cedex, France}

  \address[lpnhe]{Laboratoire de Physique Nucl\'eaire et de Hautes Energies 
    (LPNHE), Universit\'e Sorbonne, UPD, CNRS/IN2P3, 4 Place Jussieu, 75005 Paris, France }
  
  \address[skku]{Department of Electrical and Computer Engineering, Sungkyunkwan Universtity, 16419, Suwon, Gyeonggi-do, Korea}
  \address[kek]{Institute of Particle and Nuclear Studies, KEK, 1-1 Oho, Tsukuba, Ibaraki 305-0801, Japan }

  \cortext[cor_iphc]{S.\,Bilokin is now at IPHC CNRS/IN2P3 from Strasbourg (France);  M.\,Rubio-Roy is now at SPINTEC CNRS from Grenoble (France); and L.\,Lavergne is now at IRAP from Toulouse (France)}
  \cortext[cor1]{Corresponding author}

\begin{keyword}
  Calorimeter methods, calorimeters, Si and pad detectors
\end{keyword}


\begin{abstract}
The technological  prototype of the CALICE highly granular silicon-tungsten electromagnetic 
calorimeter (SiW-ECAL) was tested in a beam at DESY in 2017. 
The setup comprised seven layers of silicon sensors. Each layer comprised four sensors, with each
sensor containing an array of 256 $5.5\times5.5$ mm$^2$ silicon PIN diodes.
The four sensors covered a total area of $18\times18$ cm$^2$, and comprised a total of 1024 channels.
The readout was split into a trigger line and a charge signal line.
Key performance results for signal over noise for the two output lines are presented, together
with a study of the uniformity of the detector response. Measurements of the response to electrons for the
tungsten loaded version of the detector are also presented.
\end{abstract}

\end{frontmatter}


\section{Introduction}

The International Linear Collider (ILC) has been proposed for the next generation of
 $e^{+}e^{-}$
linear colliders \cite{Behnke:2013xla,Baer:2013cma,Adolphsen:2013jya,Adolphsen:2013kya,Behnke:2013lya}.
It will
provide collisions of polarized beams with centre-of-mass energies between 250 GeV and 1 TeV.
These collisions will be studied by multipurpose detectors.
The two proposed projects~\cite{Behnke:2013lya} --  the International
Large Detector (ILD) and the Silicon Detector
(SiD) -- are designed to exploit Particle Flow (PF) techniques  \cite{Brient:2002gh,Morgunov:2004ed}, for the
reconstruction of the final state particles.
The detectors are designed to provide single particle information.
This will enable the full power of PF analysis techniques to be exploited,
to provide accurate measurements of final state systems.
PF techniques can improve particle energy/momentum measurements by suitably
weighting calorimeter and tracker signals. This requires highly granular, compact
and hermetic calorimetry. A solenoidal magnetic field, of 3.5 - 5.0 T, is required to separate the
charged particles. The calorimeters must be placed inside the solenoidal structure
to avoid energy losses in dead material, and as a consequence the calorimeters must be compact.
Significant R\&D for calorimetry at future linear colliders has been
carried out by the CALICE collaboration.

The performance, at a test beam, of the technological prototype of the silicon-tungsten electromagnetic calorimeter, the SiW-ECAL, is reported in this document.
The SiW-ECAL is the baseline choice for the ILD ECAL. It will be placed, together with the hadronic calorimeter, inside a magnetic field of at least 3T.
Silicon constitutes the active and tungsten the passive material of the detector. 
The overall thickness is 24 radiation
lengths ($X_{0}$) or about 1 interaction length.
The baseline ILD electromagnetic calorimeter is compact, with high granularity in 3 dimensions.
In the barrel region, the calorimeter comprises 26 to 30 layers, and is contained within a radial distance of only 23 cm.

The active sensors will be segmented in squared cells of about $5\times5$ mm$^2$,
for a total of $\sim$80 million readout channels for the barrel region of the ECAL of the ILD.
The dynamic range in each channel was determined by the requirement to measure the energy deposited, at normal incidence,
by a minimum-ionizing-particle (MIP), for calibration purposes, and from the maximum particle energies expected at ILC.
The dynamic range will span 0.5 to 3000 MIPs.

The very-front-end (VFE) electronics will be embedded inside the calorimeter modules.
To maximize the hermeticity of the calorimeter, and to minimize external services, the modules will be cooled passively, through the body of the calorimeter.
To reduce the overall power consumption, "power pulsing" will be used. The electronics will be powered
during the $\sim$1-2 ms arrival duration of the  $e^{+}e^{-}$ bunch trains, and underpowered (the bias currents of the electronics will be shut down)
for $\sim200$ ms until the next cycle. The calorimeters will operate in self-trigger mode (each channel featuring a discriminator for an internal trigger decision), with on-chip zero suppression

\section{The SiW-ECAL technological prototype}

The first SiW-ECAL prototype, the SiW-ECAL physics prototype \cite{Anduze:2008hq}
was extensively tested at DESY, FNAL and CERN~\cite{Adloff:2011ha,Adloff:2008aa,Adloff:2010xj,CALICE:2011aa,Bilki:2014uep}.
The VFE was placed outside the active area with no particular constraints on power consumption.
The prototype consisted of 30 layers of silicon sensors alternated with tungsten plates 
yielding a total of 24 $X_{0}$.
The active layers were made of a matrix of $3\times3$ silicon sensors of 500 $\upmu$m thickness. Each of these sensors was segmented in matrices of
$6\times6$ squared pixels of $10\times10$ mm$^2$.

The new SiW-ECAL 'technological' prototype addresses the main technological challenges associated with the final detector: compactness,
power consumption reduction through power pulsing and embedded VFE.

\begin{figure*}[!ht]
  \centering
  \begin{tabular}{l}
  \begin{subfigure}{\textwidth}
    \centering
    \includegraphics[width=0.65\textwidth]{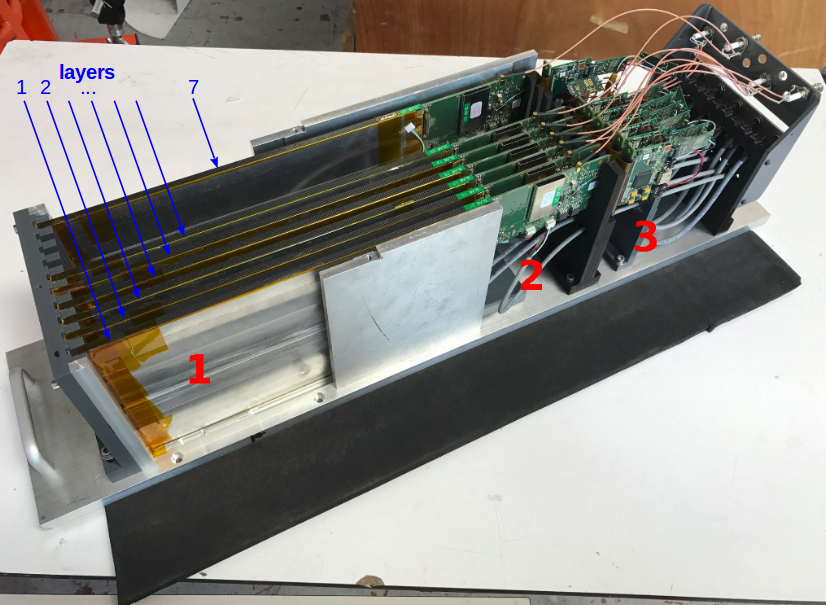} \\
    \caption{}
    \label{proto:a}
  \end{subfigure}%
  \end{tabular}
  \begin{tabular}{l}
    \begin{subfigure}{\textwidth}
    \centering
    \includegraphics[width=0.75\textwidth]{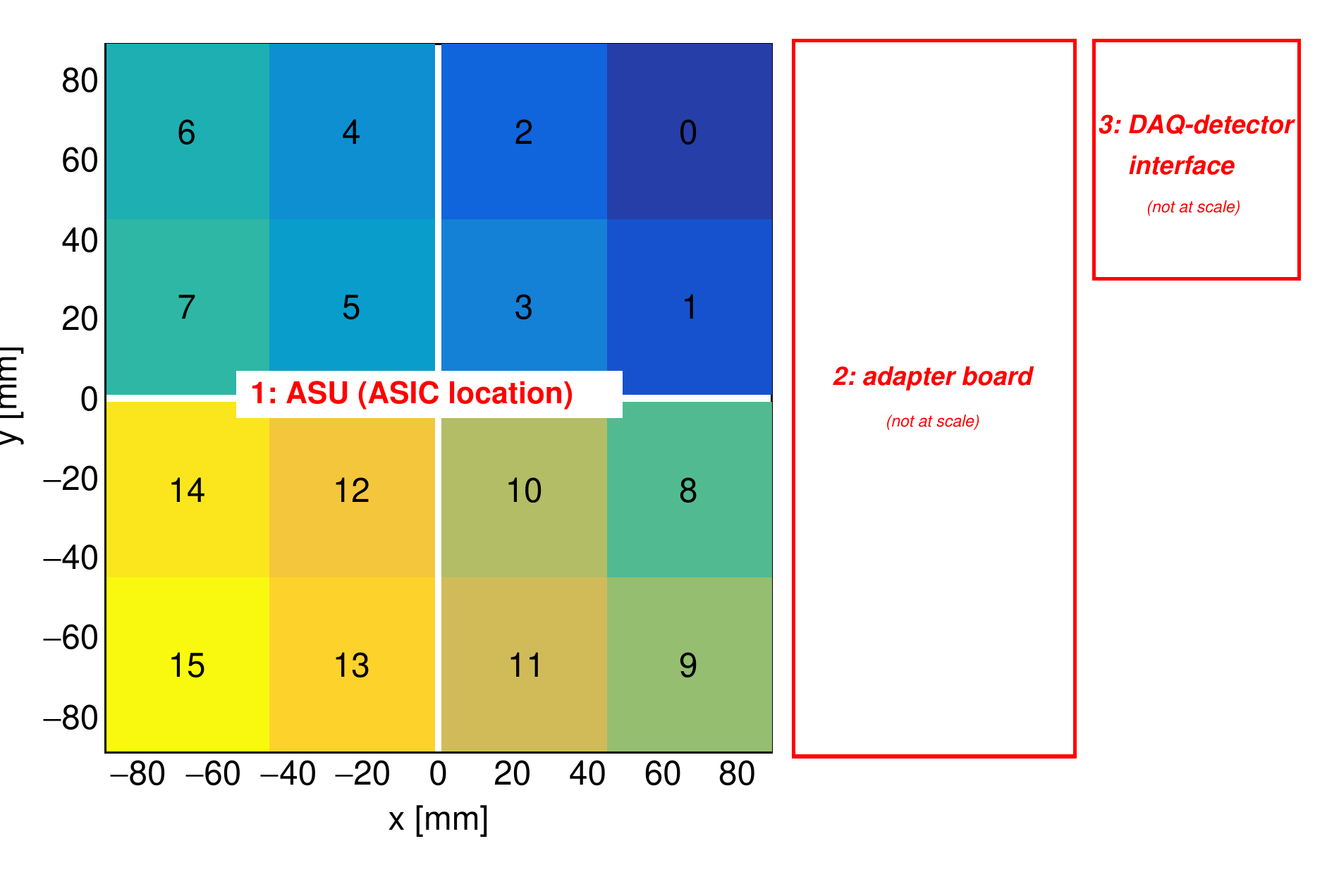} \\
    \caption{}
    \label{proto:b}
    \end{subfigure}%
  \end{tabular}
\caption{(a) Prototype with 7 layers inside the mechanical housing. The Active Signal Unit (ASU) is contained within an aluminium plate at (1). Part of the adaptor board can be seen at (2). The first component of the DAQ system is at (3).
  (b) Schematic view of the readout modules
  including the surface location readout by each ASIC, numbered from 0 to 15. Notice that the adapter board and the DAQ-detector interface drawings are not to scale.}
\label{proto}
\end{figure*}

The SiW-ECAL technological prototype setup was tested at DESY (Hamburg) in summer 2017.
The setup is shown in Figure \ref{proto:a}. It consists of 7 layers housed in a PVC and aluminium structure
that can host up to 10 layers, in slots with a 15-mm pitch. Six layers were placed in slots 1 - 6.
The 7th layer was placed in slot 10.
The layers are referenced by their numbers from 1 to 7, starting from the front face.
The detector was commissioned in the laboratory
and afterwards exposed to a low energy positron beam in the DESY test beam area (line 24) \cite{Diener:2018qap}
with the first layer located upstream .
All the results in this paper were obtained with the detector running in power pulsing mode, with
gated acquisitions of $1-3$ ms at frequencies of $1-5$ Hz.

Each layer consists of a readout module
positioned in a carbon cradle that protects the sensors
and the sides
of the module. The module is contained within two aluminium
plates to provide electromagnetic shielding and mechanical stability.
The readout module consists consists of an Active Signal Unit (ASU) and an adapter board
to a data acquisition system (DAQ)\cite{Gastaldi:2014vaa,Rubio-Roy:2017ere,Magniette:2018wdz}.
Each ASU consists of an $18\times18$ cm$^2$ sized printed circuit board (PCB).
On one side of the board is a $2\times2$ array of 4 silicon sensors.
The sensors are read out with a $4\times4$ array of 16 VFE ASICs on the other side of the PCB.
The PCB features 1024 square pads of $5.5\times5.5$ mm$^2$ which are readout by the ASICs.

The super capacitance used for the power pulsing is located on the adapter board. This
decoupling capacitor of 400 mF with 16 m$\Omega$ of equivalent serial resistance
provides enough local storage 
of power to assure stable low voltage supply during the power pulsing. The capacitor
is seen just above the red number 2 in Figure \ref{proto:a}.
A schematic drawing of a readout module
is shown in Figure \ref{proto:b}.

The VFE ASICs are 16 SKIROC2 \cite{Callier:2011zz,Amjad:2014tha,Suehara:2018mqk}
(Silicon pin Kalorimeter Integrated ReadOut Chip), 
that have been designed for the readout of silicon PIN diodes.
The SKIROC2 are enclosed in an LFBGA package and are bump bonded to the PCB.
The version of the ASUs tested in this beam test has a thickness
of 2.8 mm, including the ASICs in its current packaging.

The SKIROC2 comprises 64 readout channels. Each channel consists of a variable-gain
preamplifier followed by two branches: a fast line to provide a self-trigger decision
and a slow line for dual-gain charge measurement.
The gain on the preamplifier is set by changing the feedback capacitor.
The fast line consists of a high gain variable CRRC shaper followed by
a low offset discriminator to make the trigger decision.
A common 10-bit DAC provides the threshold of the discriminator.
The slow line is made of a low gain and a high gain CRRC
shaper to handle the extensive dynamic range.
When a channel triggers, a track and hold cell is used to record the signal at its peaking time
in the dual gain line. The levels of all the other
channels are then also recorded.
The triggers are timestamped within each gated acquisition with a slow clock
of 5 MHz. The timestamp numbers are called bunch crossing identifiers (BCID).
The charges are stored in the buffers of a Switched Capacitor Array, SCA, and later converted by a 12-bit Wilkinson ADC.
In power pulsing mode the ASIC consumes 27 $\upmu$W per channel.

The four sensors consist of $90\times90$ mm$^{2}$ silicon sensors
320$\pm15\,\upmu$m thick with high resistivity (larger than 5000 $\Omega\cdot$cm).
Each sensor is subdivided in an array of 256 PIN diodes of $5.5\times5.5$ mm$^2$
each connected to the PCB pads by a dot of conductive glue.
High voltage to deplete the sensors is delivered to the sensors via a
100 $\upmu$m copper foil isolated from the rest of the setup by a Kapton sheet.

\section{Performance at the DESY positron beam}
\label{sec:beamtest}

The beamline at DESY provides continuous positron beams in the energy range of 1 to 6 GeV with
rates from a few hundreds of Hz to a few kHz with a maximum of $\sim 3$ kHz for 2-3 GeV. 
DESY also provides access to a large bore 1 T solenoid.
The physics program of the beam test can be summarized as follows:

\begin{enumerate}
\item Calibration without tungsten absorber using 3 GeV positrons interacting, approximately, as MIPs.
  The beam was directed normally to 81 positions, equally distributed over the surface of the modules.
\item Test in a magnetic field up to 1 T using a 3 GeV positron beam.
  For this test, a plastic made structure was designed and produced to support a readout module.
\item Response to electrons of different energies with a fully equipped detector, comprising detector layers {\it and} tungsten absorber. 
\end{enumerate}

The commissioning of the prototype is discussed in Section \ref{sec:commissioning}. 
The results of the pedestal, noise and MIP calibration in Section \ref{sec:calib}.
The performance in a magnetic field is discussed in Section \ref{sec:magnetic}.
The response of the detector to electromagnetic shower events is discussed in Section \ref{sec:showers}.

\subsection{Commissioning of the detector}
\label{sec:commissioning}

Earlier experience with the SKIROC2 ASIC has been reported in \cite{Amjad:2014tha,Suehara:2018mqk,Balagura:2017pka}. 
For the following, the internal SKIROC2 parameters determined in these references are adopted
except if otherwise stated. It has been estimated that
a MIP traversing the PIN perpendicular to its surface produces, on average,
a signal of 4 fC.
Since most of the data taking program consisted of the recording
of MIP level signals, we set a low value of the feedback capacitance of the preamplifier to $1.2$ pF  to obtain
gains of $71.25\,{\rm mV/fC}$ and $7.125\,{\rm mV/fC}$ in the dual-gain charge measurement.
At high gain, the SKIROC2 has a linearity of better than 90\% for signals in the range of 0.5-200 MIPs.
This is sufficient for the analysis of electromagnetic showers created by positrons of a few GeV. All data presented here were taken at high gain.

\subsubsection{Masking of noisy readout channels}
\label{sec:noisy}

At the start of commissioning, channels were disabled if they caused triggers
above a noise threshold of 1 MIP \cite{Bilokin:2018gfn}. The disabled channels can be classified into four categories:

\begin{itemize}
\item Routing issues: The same channels trigger in all layers. This hints at a issue with the routing of the PCB. 
  Following a conservative approach, we added to this set all channels that were noisy in at least three modules.
\item ASIC issues: In case 70\% of the channels of an ASIC triggered, the entire ASIC was disabled.
  \item Sensor issues: If a wafer showed problems e.g. like high leakage currents, all channels connected to the wafer were disabled.  
\item Others.
\end{itemize}

\begin{figure}[h!]
  \includegraphics[width=0.5\textwidth]{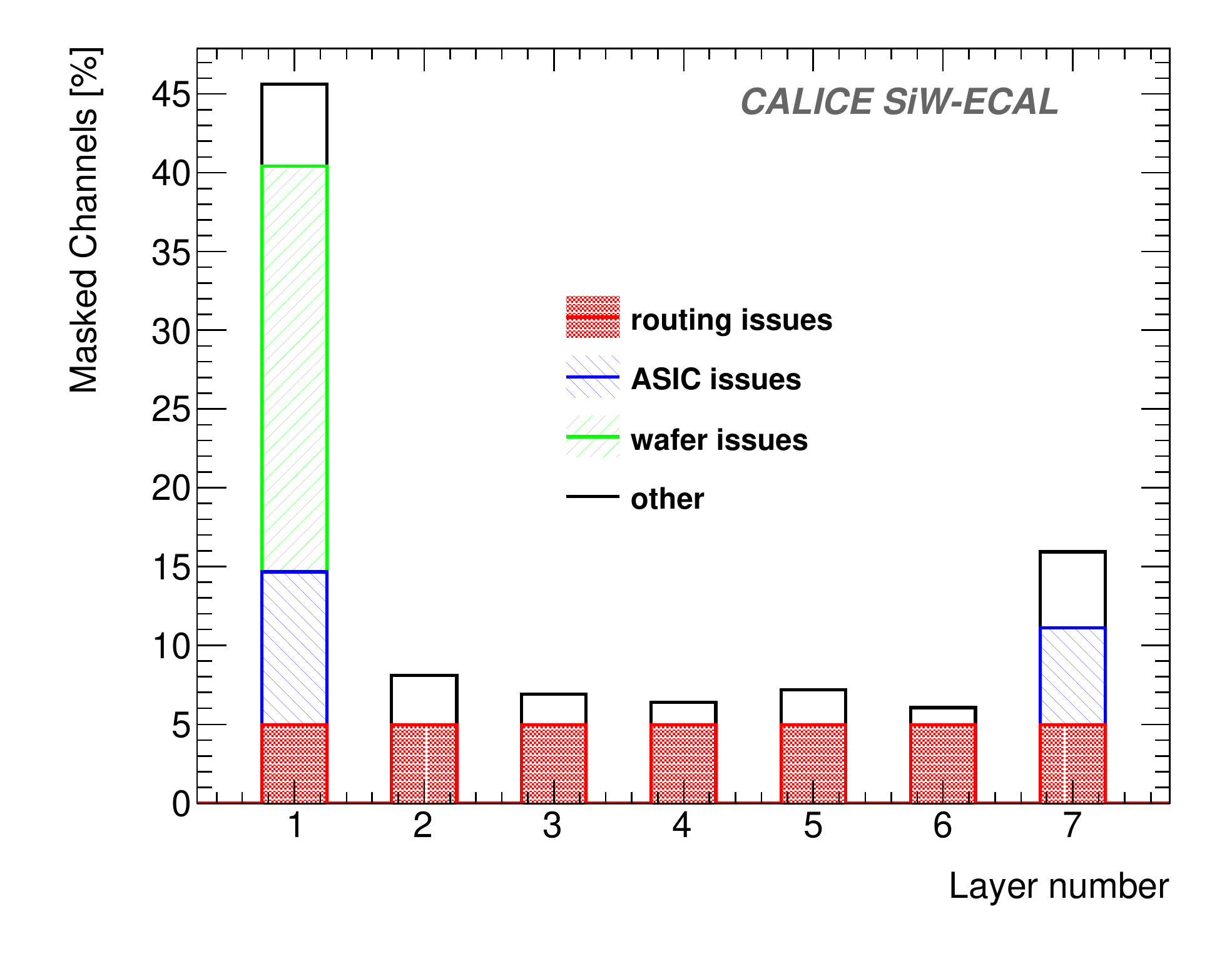} 
\caption{Fraction of channels with counts above one MIP in noise runs and which have been disabled for the data taking.}
\label{noisycells}
\end{figure}

The results of this study are summarised in Figure \ref{noisycells}.
All these channels have been disabled during the data taking.

\subsubsection{Trigger threshold optimization and signal-over-noise ratio for the trigger}
\label{sec:comm_trigger}

For each ASIC, the threshold of the internal trigger
was optimized through dedicated scans of trigger threshold values
with noise signals or with injected signals of different amplitudes.
The scans start from high values of the thresholds, which are successively
lowered while the number of signals recorded by each channel is counted. 
The resulting curves can be approximated by an inverted error function.
The optimal threshold was chosen as the threshold at 50\% plus 3 times the
width of the curve. The width was defined as half the difference between
the thresholds at 84\% (50+34\%) and 16\% (50-34\%).
The average of the optimal thresholds for all channels
is used as the global optimal threshold to be used by the ASIC.
If the analysis failed in more than 50\% of channels ({\it i.e.} because they are masked),
an {\it ad-hoc} high value of the threshold is chosen for the ASIC (250 in DAC units).

The width of the threshold curve depends on the ratio
of the frequency of the white noise of the electronics to the clock speed of the readout.
Therefore, the width of the threshold curve does not provide a direct measure
of detector noise.
Alternatively, the scan can be repeated with external signals, provided either by beam particles or by electronic injection.
Following the second approach, amplitudes equivalent to 1 and 2
MIPs were injected into channels of an ASIC mounted on a test board.

\begin{figure}[h!t]
    \centering
  \begin{tabular}{l}
	\includegraphics[width=0.5\textwidth]{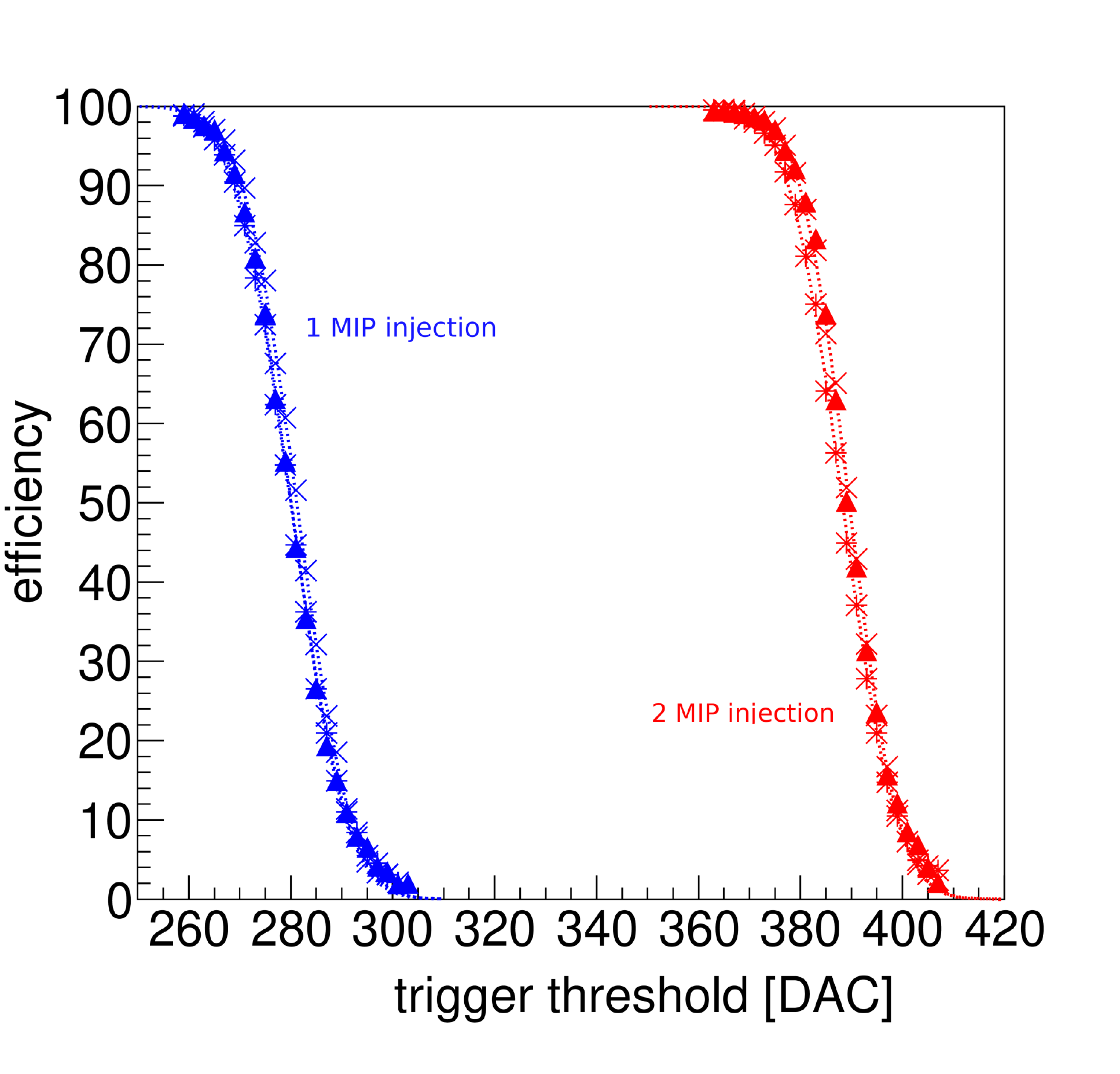}\\ 
	\end{tabular}
  \caption{Two sets of threshold curves. Each set consists of threshold curves for individual channels in which a charge is artificially injected using a pulse generator.
    For the set in the left (in blue), an equivalent of 1 MIP was injected. For the set in the right (red), a equivalent of 2 MIPs was injected.
    These curves were obtained in a dedicated test board \cite{Suehara:2018mqk} to test single SKIROC2 ASICs.}
\label{scurves_injection}
\end{figure}

The result of these scans is shown in Figure \ref{scurves_injection} for several channels of one SKIROC.
These curves are used to establish the trigger threshold and to estimate
the trigger signal-to-noise, $(S/N)_{trigger}$. The signal-to-noise is defined as the ratio between the distance of the
1 MIP and 2 MIP threshold curves scans, at 50\% efficiency, to the width of the 1 MIP threshold scan curve.
From the measurement shown in Figure \ref{scurves_injection}, we extract a $(S/N)_{trigger}$ ratio of about 12.8.

\begin{figure}[h!t]
  \centering
  \includegraphics[width=0.5\textwidth]{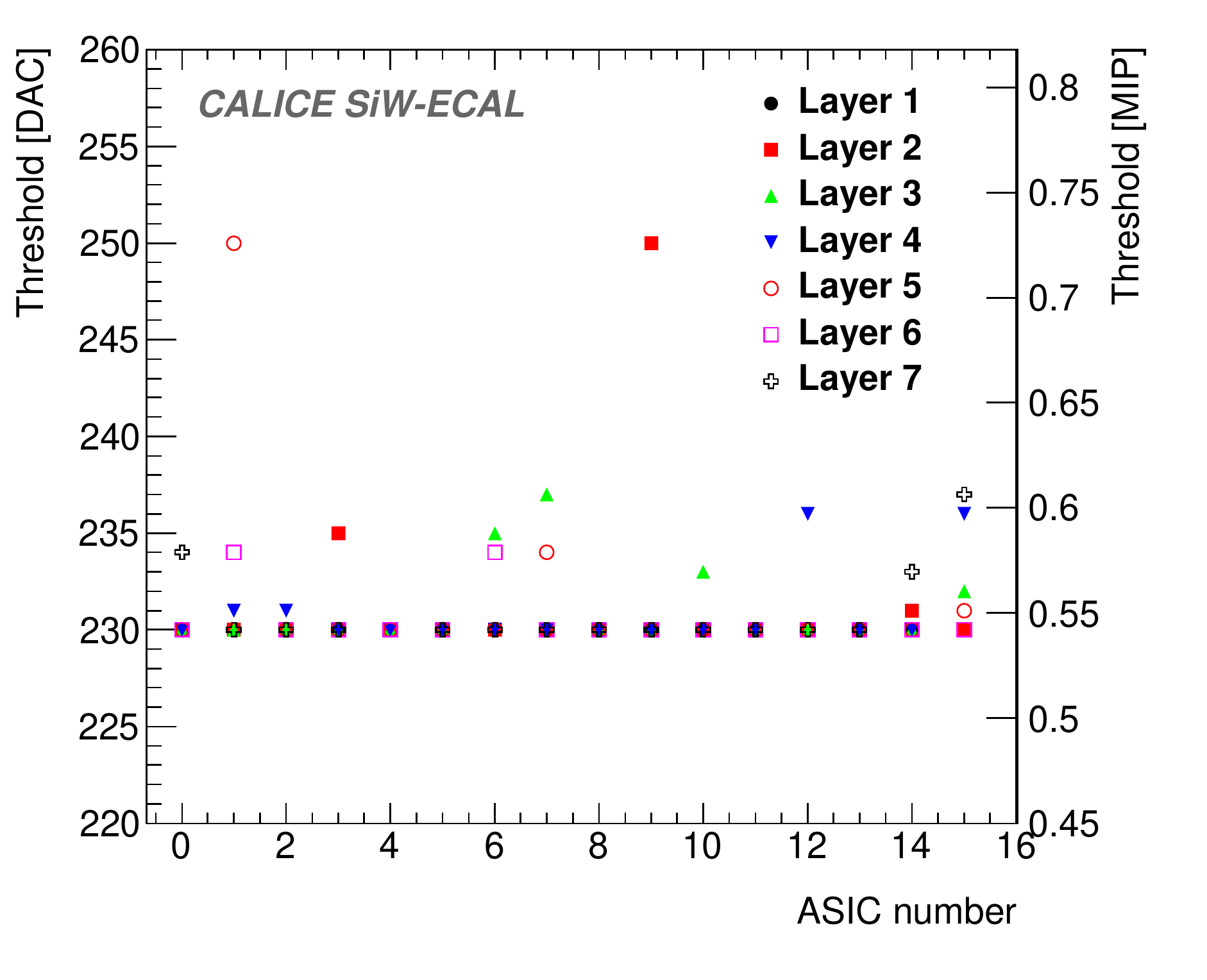}
  \caption{Summary of the trigger threshold settings in the internal DAC units (left vertical scale) and in MIP units (right vertical scale).}
\label{trigger_thresholds}
\end{figure}

The trigger threshold calibration (from DAC units to MIPs) is extracted from these curves following a simple formula:

\begin{equation}
	\begin{split}
1 -Threshold[MIP] = \\
\frac{DAC_{50\%}(1\,MIP)-Threshold[DAC]}{DAC_{50\%}(2\,MIP)-DAC_{50\%}(1\,MIP)}
	\end{split}
\end{equation} 

where $DAC_{50\%}(X\,MIP)$ stands for the DAC value at the $50\%$ point of the threshold scan curve obtained with $X$ MIP signals.
This calibration has been applied to all readout modules: see Figure \ref{trigger_thresholds}.
Further dedicated data taking runs with particle beams are needed in the future to measure the systematic errors
and spread in performance, between
SKIROCs and modules, from the calibration data.


\subsection{Response to 3 GeV positrons acting as MIPs}
\label{sec:calib}

All the analyses presented here rely on the selection and identification
of track like events. The first step of the reconstruction is the rejection
of two types of fake events due to the ASIC
design: {\it a)} triggers in consecutive BCIDs
due to the preamplifiers sensitivity to instabilities of the power supply; {\it b)}
an artefact of the event validation logical sequence of the ASIC. These fake
events are removed offline using timing information.
Electromagnetic showers are rejected by requiring
less than 5 triggers per module.
Tracks are reconstructed from the seven layers by selecting events
where the timestamps from the layers are clustered together, allowing for a
tolerance on the BCID of $\pm1$. An event is accepted if at least
three of the layers were clustered together in a track traversing the modules in the
same position
for all modules with a tolerance of $\pm5.5$ {\rm mm} per layer.

\subsubsection{Pedestal and noise determination}
\label{sec:pedestal}

\begin{figure}[h!t]
  \centering
  \begin{tabular}{l}
    \includegraphics[width=0.5\textwidth]{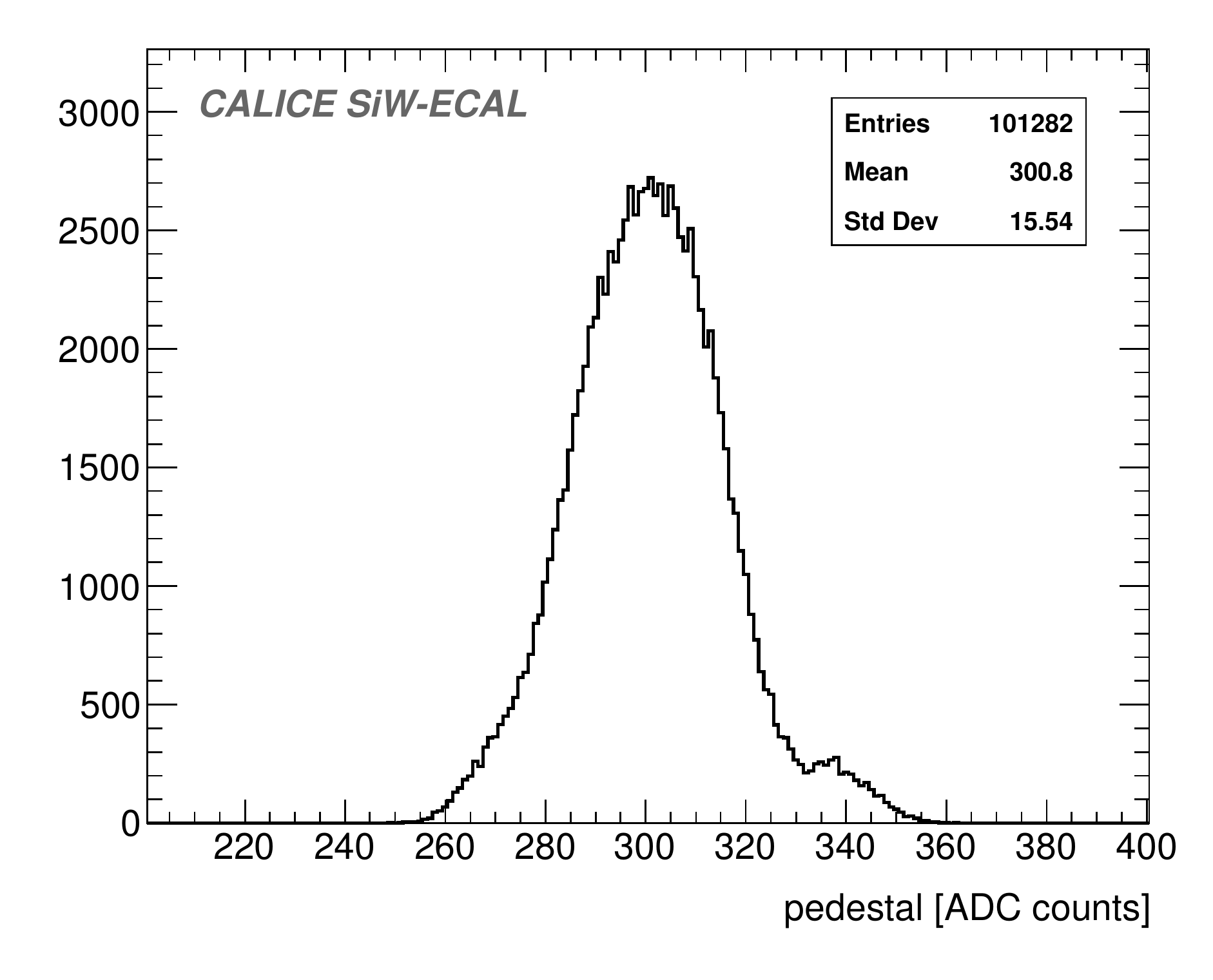} \\
    \includegraphics[width=0.5\textwidth]{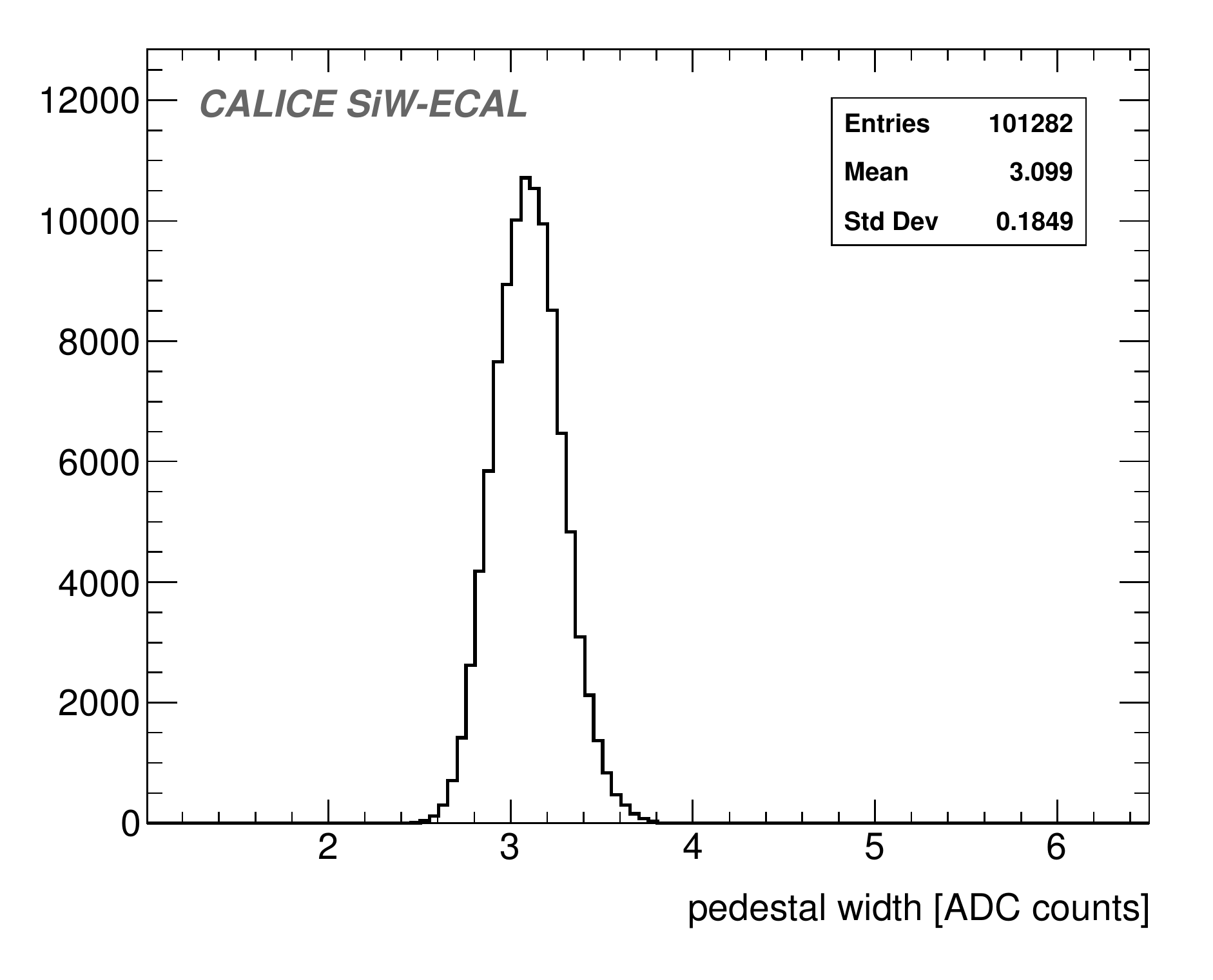}
  \end{tabular}
  \caption{
  Distribution of the pedestal mean positions (upper) and widths (lower), from the high gain charge measurements,
for Gaussian fits of the pedestal data, for all individual channels and SCAs in the setup.
}
\label{pedestal_all}
\end{figure}

The pedestal positions and widths, from the high gain charge measurement line,
are shown in Figure \ref{pedestal_all}.
Masked channels, or channels involved in the trigger region, were not included. A Gaussian fit was made to the ADC distribution from each channel.
The pedestal position was taken from the mean of the Gaussian. The width of the pedestal was taken from the Gaussian standard deviation.
These values were calculated for all ASICS, channels and SCA under test. These data are used to perform a pedestal subtraction to the
raw ADC signal. This process is done on a layer, ASIC, channel and SCA basis.

The calculated pedestal positions and widths of the ADC for all non-triggered and non-masked
channels and SCAs are shown in Figure \ref{pedestal_all}.
The non-gaussian spread of the pedestal position is explained by the fact that each SCA has its own pedestal value. More important is that the 
width of the distributions is very similar for all
channels and SCAs.
This shows the good level of homogeneity of the noise levels across the system.

\subsubsection{MIP calibration}
\label{sec:mip}

For all channels under test, the resulting charge spectrum of the triggered channels
after pedestal correction is fit by a Landau function convoluted with a Gaussian if the number of events was
larger than 1000. 
The most-probable-value of the Landau function is taken as the MIP value, allowing thus for a direct
conversion from ADC counts to energy in MIP units.

\begin{figure}[h!t]
  \centering
  \includegraphics[width=0.5\textwidth]{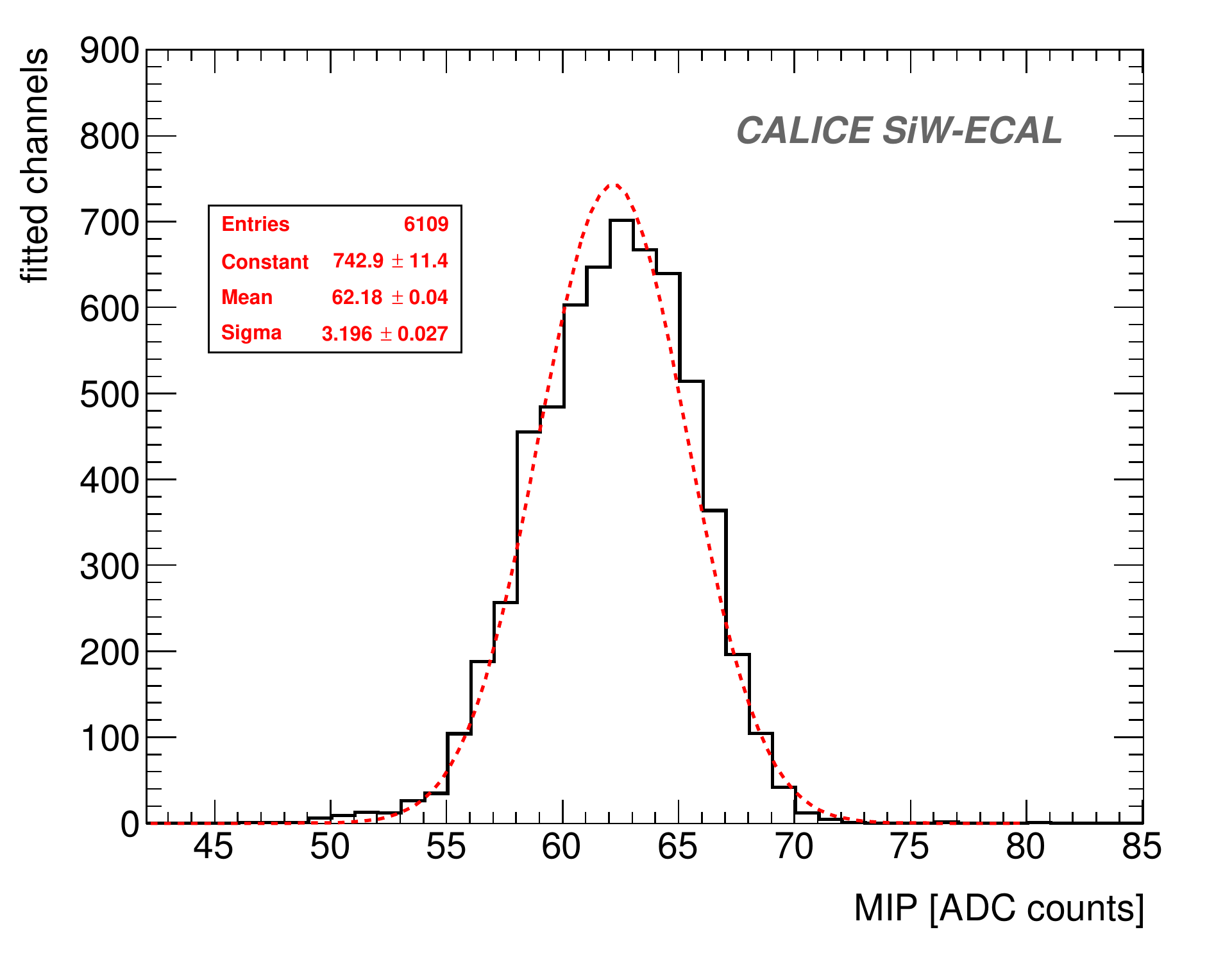} \\
\caption{The sensor MIP calibration, for all calibrated channels, with the signal yield in ADC counts.}
\label{mip}
\end{figure}

The fit succeeded in 98\% of the cases. Figure \ref{mip} shows the results of the MIP calibration for the successful fits.
The calibration shows that the response to MIPS across the detector is the same to within 5\%.

\subsubsection{Signal-over-noise ratio for the charge measurement of triggered channels}
\label{sec:sn}

The signal over-noise-ratio for the charge measurement of triggered channels, $(S/N)_{charge}$, is defined 
as the ratio of the calculated MIP value to the noise (the pedestal width) for each channel and SCA.
The measured value, for all channels under test, is:
\begin{equation}
(S/N)_{charge}=20.4\pm1.5
\end{equation}

\begin{figure}[h!t]
  \centering
  \includegraphics[width=0.5\textwidth]{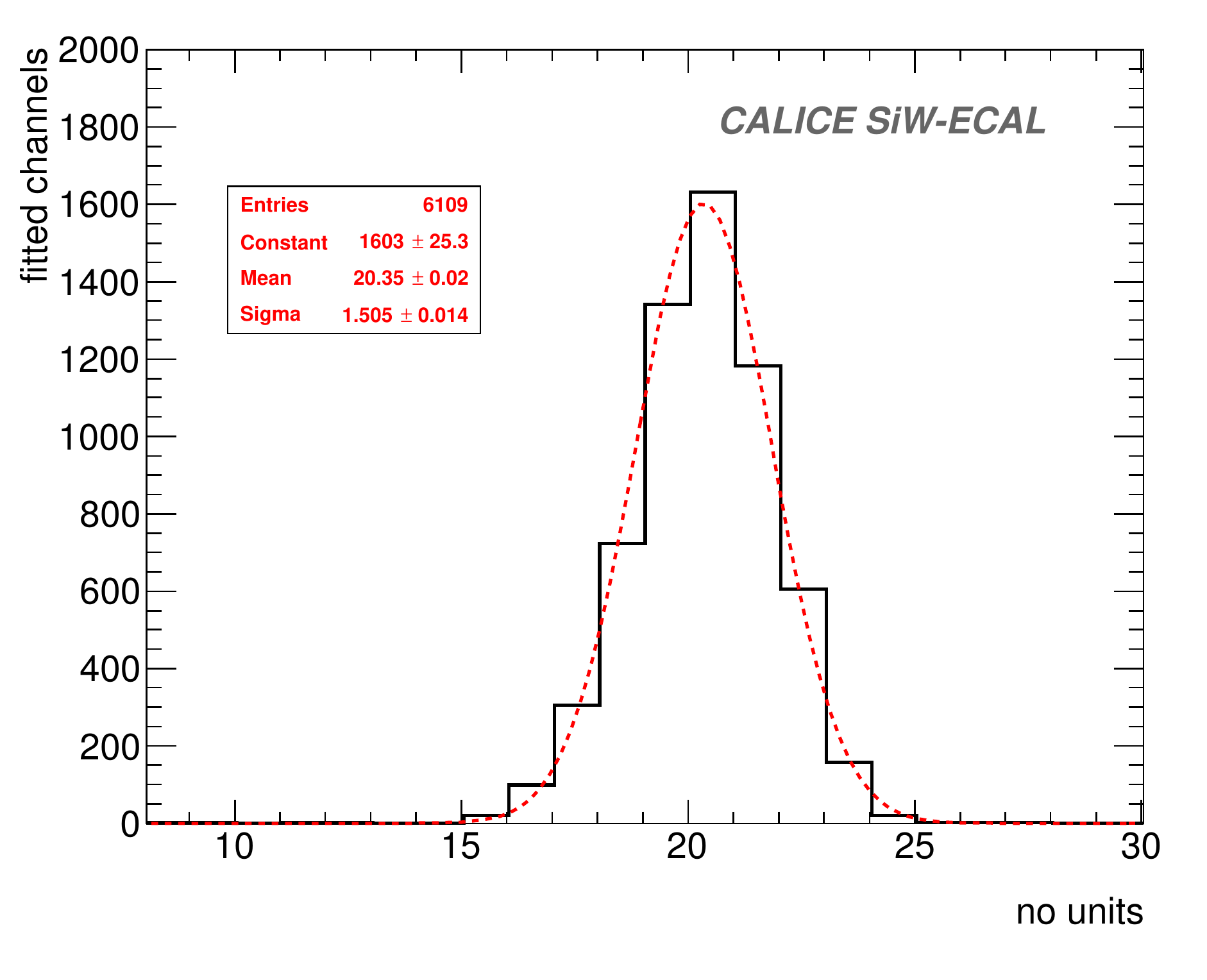}  
\caption{Distribution of the $(S/N)_{charge}$ for all calibrated channels.}
\label{mipandSN}
\end{figure}

The results are summarized in Figure \ref{mipandSN}.
The excellent ratio ensures that low energetic hits just above the trigger threshold can actually be used for the event reconstruction.

\subsubsection{MIP detection efficiency}
\label{sec:mip}

\begin{figure}[!t]
  \centering 
  \includegraphics[width=0.5\textwidth]{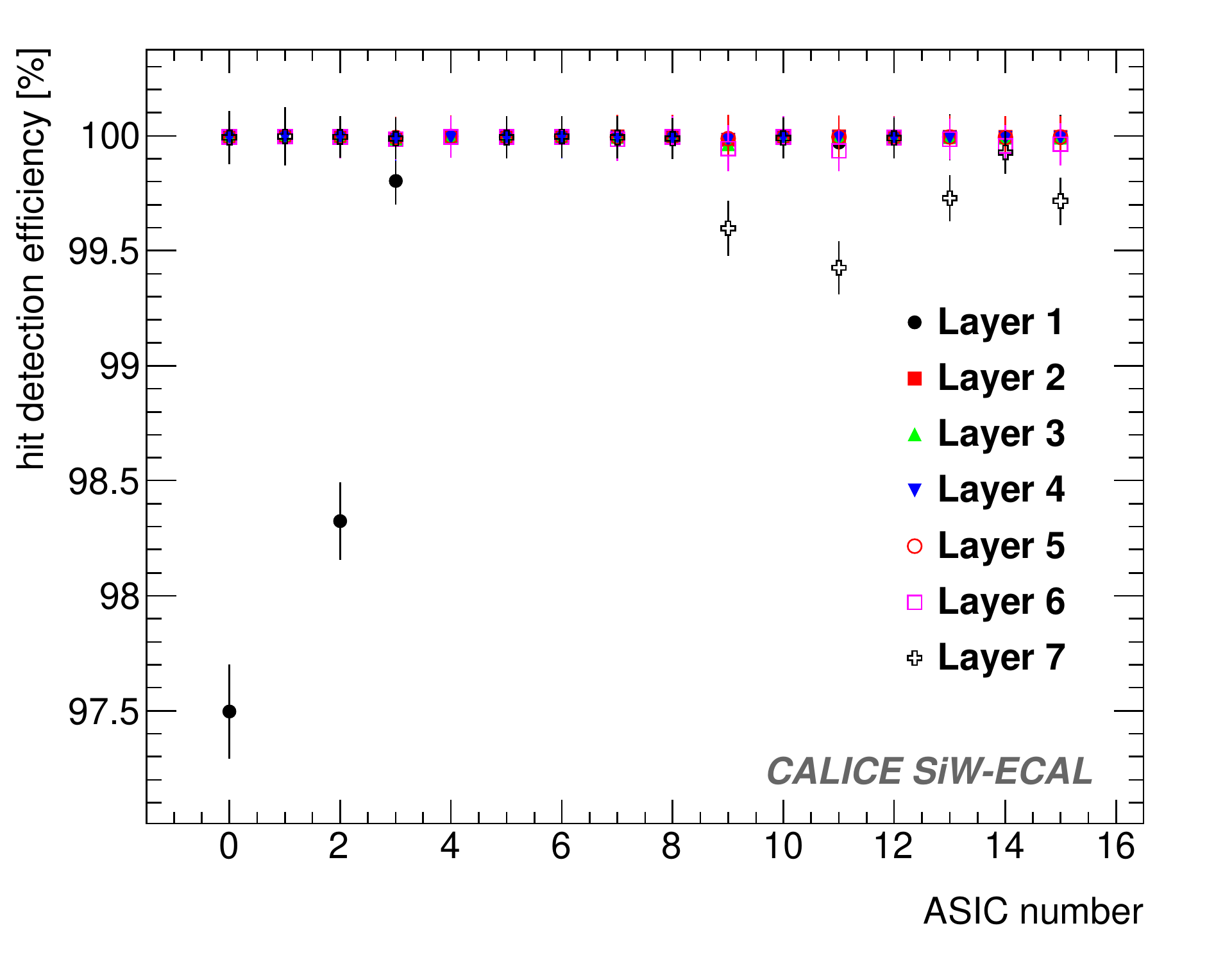} \\
  \caption{The MIP detection efficiency, for high purity track events, using
through-going positrons (no tungsten absorber present). The average for all readout channels in each ASIC is shown. The error bars correspond to the standard deviation of the efficiency distribution of all channels in the ASIC.}
\label{efficiency}
\end{figure}

To evaluate 
the single hit detection efficiency, we increased the purity of our track sample
by requiring tracks with at least 4 layers, with the hits corresponding to the same beam location.
For the remaining three layers, the search region is extended to the eight channels
surrounding the target channel. Residual spurious signals are filtered by requiring a minimum energy deposition of 0.3 MIP.
The average efficiency, for every ASIC, is shown in Figure \ref{efficiency}.
Except for a few exceptions, the efficiency is 
compatible with $100\%$.
Low efficiencies in the first layer are related to the presence of
noisy channels not spotted during the commissioning. These channels
saturate the DAQ in their ASICs earlier than in others.
In the last layer, we also observe a few small deviations
which are associated with channels in the periphery, suggesting a small misalignment of the last layer.

\subsection{Performance of a single readout module in a magnetic field}
\label{sec:magnetic}

The study of the performance of the module inside
magnetic fields was carried out in four consecutive steps:
a reference run without magnetic field; a run with a magnetic field of 1 T; a second run with 0.5 T; and a final run repeated without magnetic field.
The beam, 3 GeV positrons, was incident on the area of the PCB readout by the ASIC number 12.
During all this time, the readout module was continuously monitored and did not
show any technical issue or change of performance.

\begin{figure}[!t]
  \centering
  \begin{tabular}{l}
    \includegraphics[width=0.5\textwidth]{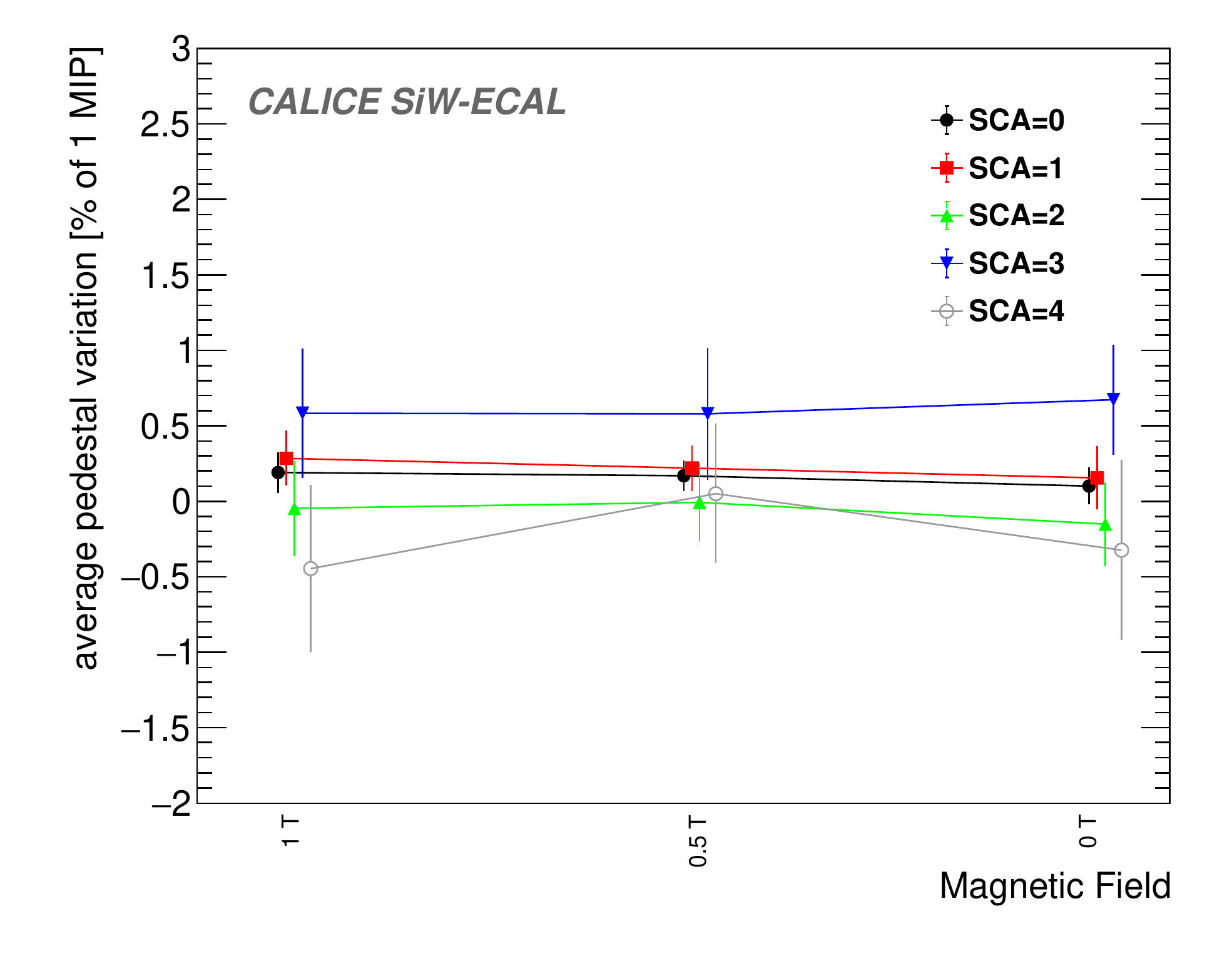} \\
    \includegraphics[width=0.5\textwidth]{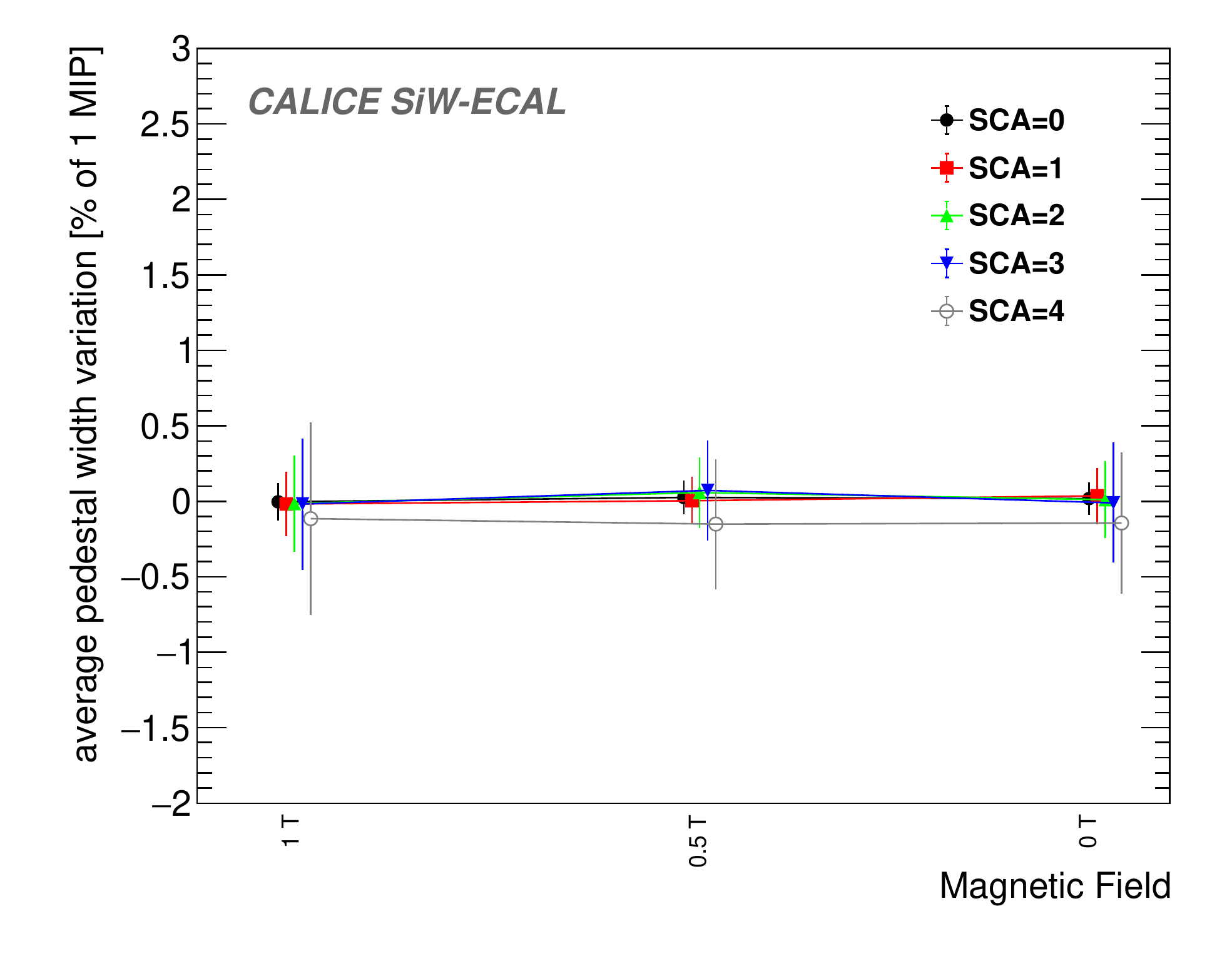}
  \end{tabular}
  \caption{
    Average deviation of the pedestal mean position (up) and width (down) during the different runs at different magnetic field (1T, 0.5 T and 0T)
    compared with data taken outside of the magnet structure. The error bars correspond to one standard deviation of the averaged values.
    The results for several SCA is presented in each plot, with a slight shift in the $x$-axis to artificially separate the points for better visualization.
  }
\label{pedestal_magnetic}
\end{figure}

Due to the lower beam rates in this beam area at DESY (and the technical character of the test),
we did not collect enough data to present a reliable comparison with the calibration described in the previous sections.
However, a detailed study was made of the stability of the pedestal position and width of all channels
in ASIC 12. Their values are well compatible with the values obtained in Section \ref{sec:calib}, as is shown in Figure \ref{pedestal_magnetic}.
We see that the agreement is perfect within the statistical uncertainties.
Due to the lower rates in this beam area, the
analysis was only carried out on a few channels and memory cells.

\subsection{Performance of the SiW-ECAL for low energy electromagnetic showers.}
\label{sec:showers}

For the last part of the beam test, tungsten sheets with a thickness of 1.2 $X_{0}$ each were inserted in front of each of the seven layers yielding a total depth of 8.4 $X_{0}$.


The beam was directed in the centre of the region covered by ASICs 12-15, see  Figure \ref{proto}. 
Between 5000 and 10000 triggers were recorded for electron beam energies of 1 to 5 GeV.
For electrons at 5.8 GeV, $\sim$1000 triggers were recorded.
The analysis was carried out using procedures described in section \ref{sec:calib}.
Events were selected by timestamp. Single cell calibrations were
applied. 
In the present analysis, no constraint on the number of triggers per 
layer was applied.

\begin{figure}[!t]
  \centering
  \includegraphics[width=0.5\textwidth]{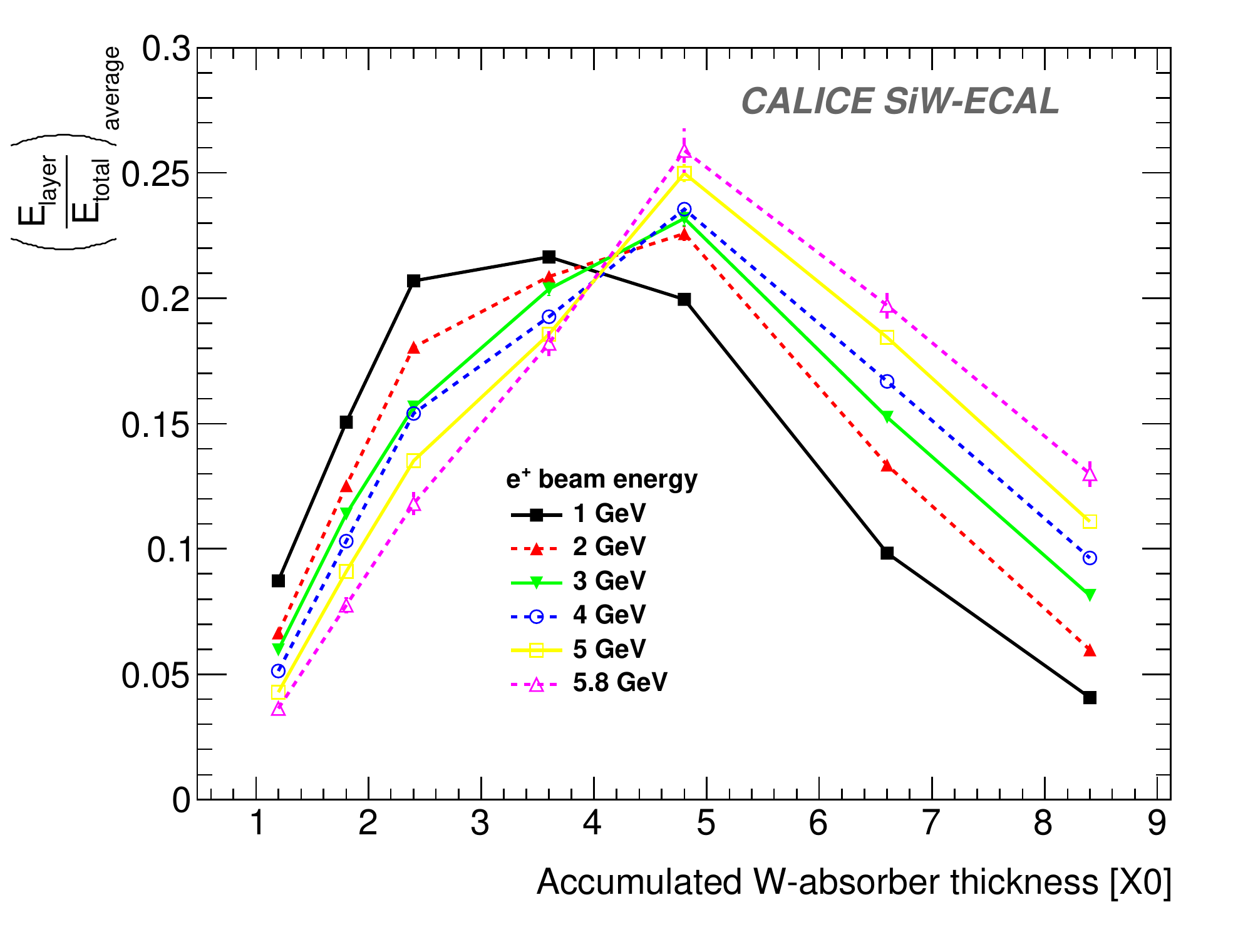}\\
  \caption{Electromagnetic shower profiles for different beam energies. In the x-axis, we show the accumulated thickness of tungsten absorber in front
    of every active layer. In the y-axis, the averaged fraction of the total energy measured by every active layer in each reconstructed event. }
\label{showers}
\end{figure}

Figure \ref{showers} shows the average energy measured per layer as a function of the amount of tungsten absorber in each layer, in $X_{0}$ units.
The profile of the energies, measured across the layers, is as expected for electromagnetic cascades
generated mainly in the tungsten absorber.
The measured distributions for the various energies behave qualitatively as expected. 
The shower maximum is at around 5 $X_{0}$ and shifted towards smaller $X_{0}$ for the smallest energies under study. 
Finally one observes that the longitudinal leakage increases with increasing energy.
A more detailed study of the detector response to electromagnetic showers including comparisons to shower simulation programs like GEANT4 is left for a future publication.

\section{Summary and outlook}
\label{sec:summary}

In this document, we present the results on the commissioning 
in beams of a small but fully equipped SiW-ECAL technological prototype made of detection units 
featuring the main characteristics foreseen for future high energy $e^{+}e^{-}$ collider detectors. 
The prototype comprised seven independent layers of silicon sensors. Each layer had 1024 readout channels,
with embedded electronics that were designed for the ILD detector for the International Linear Collider.
The prototype was tested in a setup optimised for a physics program dedicated to the study of electromagnetic showers. 
The layers were entirely operated in power-pulsed mode.

Using through-going positrons as MIPs, the beam test has allowed for an in-depth evaluation of the performance of the detector. 
The response to MIPs is homogeneous, with 98\% of the unmasked channels calibrated within a spread of 5\%.
The trigger line signal-over-noise ratio, defined as the ratio of the most-probable MIP response to the pedestal standard deviation, 
is $\sim$12.8. 
With this trigger signal-to-noise, triggers can be formed with high efficiency from signals in individual cells that are well below the MIP level. 
This will be required for effective particle flow analyses. 
New dedicated beam tests are required 
to characterise a broader set of modules and ASICS.
The MIP track detection efficiency is ~100\% for most of the channels.
The signal-over-noise ratio for the high gain charge line measurement is $\sim$20.4, for triggered channels. 
This will ensure that all channels with a trigger just above noise level can be used for data analysis.

When interleaved with tungsten plates 
the layers qualitatively exhibit the expected electromagnetic shower profiles.
Further studies and comparisons with simulations must be conducted to assess these observations quantitatively. 
These data will be complemented by subsequent beam tests, including further R\&D developments 
and a potentially larger stack with up to $\sim$20 to 30 fully equipped layers, 
and a homogeneous distribution of tungsten absorber amounting up to $\sim$24 $X_{0}$.

\section*{Acknowledgments}

This project has received funding from the European Union{\textquotesingle}s Horizon 2020 Research and Innovation program under Grant Agreement no. 654168.
This work was supported by the P2IO LabEx (ANR-10-LABX-0038), excellence project HIGHTEC,
in the framework {\textquotesingle}Investissements d{\textquotesingle}Avenir{\textquotesingle}
(ANR-11-IDEX-0003-01) managed by the French National Research Agency (ANR).
The research leading to these results has received funding from the People Programme (Marie
Curie Actions) of the European Union{\textquotesingle}s Seventh Framework Programme (FP7/2007-2013)
under REA grant agreement, PCOFUND-GA-2013-609102, through the PRESTIGE
programme coordinated by Campus France.
The measurements leading to these results have been performed at the Test Beam Facility at DESY Hamburg (Germany), a member of the Helmholtz Association (HGF).

\section*{References}
\bibliographystyle{elsarticle-num}
\bibliography{../../references}

\end{document}